\documentclass[aps,preprint,amsmath,amssymb]{revtex4}
\usepackage{amssymb}
\usepackage{mathrsfs}
\usepackage{graphicx}
\usepackage{subfigure}
\usepackage{hyperref}
\usepackage{dcolumn}
\usepackage{bm}
\usepackage{txfonts}

\newcommand{\mycite}[1]{\scalebox{1.0}[1.0]{\raisebox{0.1ex}{\cite{#1}}}}
\begin{document}
\title{Properties of single-layer graphene doped by nitrogen with different concentrations}
\author{Zongguo Wang}
\affiliation{
State Key Laboratory of Theoretical Physics,
Institute of Theoretical Physics,
Chinese Academy of Science, Beijing 100190, People's Republic of China
} \email{wangzg@itp.ac.cn}
\author{Shaojing Qin}
\affiliation{
State Key Laboratory of Theoretical Physics,
Institute of Theoretical Physics,
Chinese Academy of Science, Beijing 100190, People's Republic of China
}
\author{Chuilin Wang}
\affiliation{China Center of Advanced Science and Technology,
P. 0. Box 8730, Beijing 100190, China}

\begin{abstract}
Graphene has vast promising applications on the nanoelectronics and spintronics because of its
unique magnetic and electronic properties. Making use of an ab initio spin-polarized density
functional theory, implemented by the method of Heyd-Scuseria-Ernzerhof 06(HSE06) hybrid
functional, the properties of nitrogen substitutional dopants in semi-metal
monolayer graphene were investigated. We found from our calculation, that introducing
nitrogen doping would possibly break energy degeneracy with respect to spin (spin symmetry breaking)
at some doping concentrations with proper dopant configurations. The spin symmetry
breaking would cause spin-polarized effects, which induce magnetic response in graphene.
This paper systematically analyzed the dependence of magnetic moments and band gaps in
graphene on doping concentrations of nitrogen atoms, as well as dopant configurations.
\end{abstract}
\maketitle

\section{Introduction}
Since the discovery of graphene, a monolayer of carbons distributed in a honeycomb
lattice\cite{Novoselov2004}, it has attracted the most attention of
experimental\cite{Latil2007,Malard2007,Riedl2007} and theoretical\cite{Sakhaee-Pour2008,Mattausch2007,Hwang2007,Xiao2007}
research due to its novel electronic properties. Graphene is the basic structure of all
graphitic forms of carbon\cite{Geim2007}, and its unique two dimensional physical and chemical
properties make graphene potential candidate for electronic device\cite{Levy2010}, field effect
transistors(FET)\cite{Xia2010}, supercapacitors\cite{Zhang2010}, and fuel cells\cite{Kauffman2010,Seger2009,Qu2010,Shao2010}.
Since material properties are related to its structures,
substitutional doping is a powerful way to modify material properties,
and it might be expected to have fundamentally different
consequences. Chemical doping with B and N atoms is considered an effective method
to improve the electronic properties of carbon
materials, and forming $p$- and $n$-type carbon materials. It is more useful to get
$n$-type graphene in comparison with the easily obtainable
$p$-type graphene by adsorbates.
Among various doping elements, nitrogen is the most promising candidate of doping elements
in regulating the electronic properties of graphene,
due to its similar atomic size, and its five valence electrons being easily
formed strong covalence bonds with carbon atoms.

Experimental results\cite{Geng2011,Zhao2011} have reported the various structure
species of graphene with nitrogen dopants, and it has been shown that nitrogen is doped
into the graphene structure in the form of graphitic dopants. Two important factors related to
application of nanoelectronics and spintronics for graphene are the magnetic moments
and the band gaps. In order to use graphene vastly in nanoelectronics, spintronics,
and photovoltaic devices, it is important to study and exhibit electronic structures
of modified systems. The research group in Georgia Tech. had announced
that they successfully created the planar FET of graphene, and observed quantum
interference effect, and revealed excellent transport properties in specified
configurations of single graphene layers\cite{Quhe2012}. Graphene with planar
structures will be the original materials for advanced industry.
Many papers have been reported the effects of nitrogen doped graphene, but it
is still unclear that how does nitrogen doping affect the magnetism and band gaps.
Besides concentration, are there any other factors that affect the magnetism
and band structure? Inspired by these questions, employing the spin-polarized density
functional theory(DFT), we systematically analyzed the dependence of magnetic
moments in graphene on doping concentrations of nitrogen atoms, as well as dopant
configurations. We found from our calculation, that introducing nitrogen doping at
some doping concentrations with proper dopant configurations will break sublattice
symmetry, which thereafter would possibly break energy degeneracy with respect
to spin (spin symmetry breaking). The spin symmetry breaking would cause
spin-polarized effects, which induce magnetic response in graphene.
\section{Computation methods and models}
The calculations are performed by using the Vienna Ab initio Simulation Package
(VASP)\cite{Wien1994,Kresse1996} with the spin-polarized DFT implemented.
The generalized gradient approximation(GGA) in Perdew-Burke-Ernzerhof(PBE)
form\cite{Perdew1996} is employed to calculate the exchange-correlation energy.
The electron-ion interactions are described by the projector augmented wave
(PAW)\cite{P.E.Blochl1994}. The Brillouin zone is sampled by a k-point mesh
generated according to the Monkhorst-Pack scheme\cite{Monkhorst1976}. The dense
k-point meshes 4x4x1 and 7x7x1 are used for the geometry optimizations and static
structure energy calculation, respectively. A plane wave basis set is used and the
kinetic-energy cutoff is taken as 400eV. The criterion of energy convergence is
chosen to be less than $10^{-5}$eV/atom. Heyd-Scuseria-Ernzerhof 06 (HSE06) hybrid
functional has been shown as an accurate method to calculate band gaps for many
semiconductors\cite{Hummer2009} including the graphene-related $\pi$ sysytem\cite{Hod2008}.
And it also provided lattice constants and local spin magnetic moments that were in good
agreement with the experiment\cite{Marsman2008}.
The cell shape was almost kept in its original form with its
lengths and angels slightly changed during the calculations, and the atoms relaxed nearly
in a plane with a very small negligible torsion of angle out of plane around C-N bonds.

\begin{figure}
\center
\setlength{\abovecaptionskip}{-10pt}
\setlength{\belowcaptionskip}{0pt}
\includegraphics[width=3.0in]{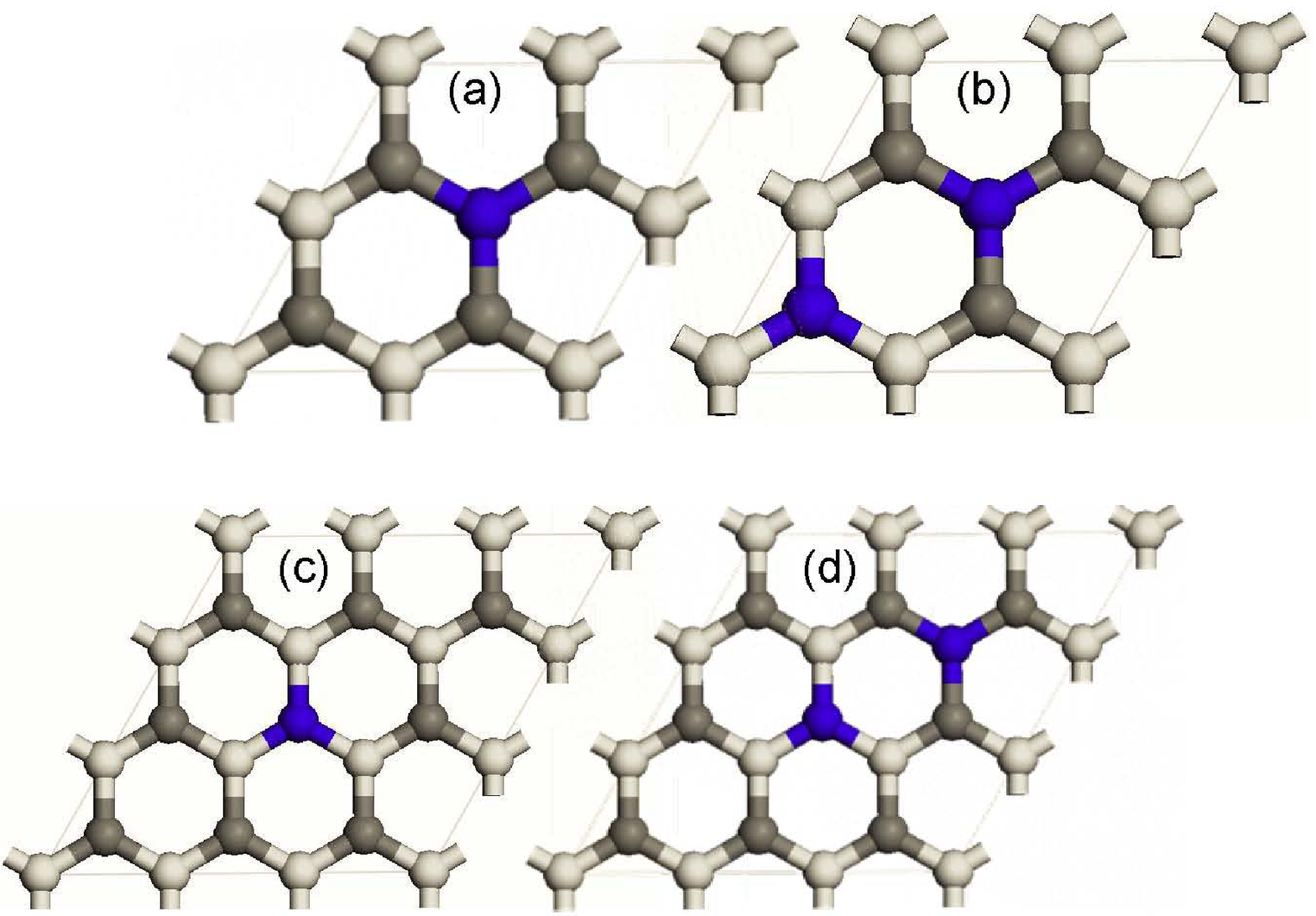}
\includegraphics[width=3.2in]{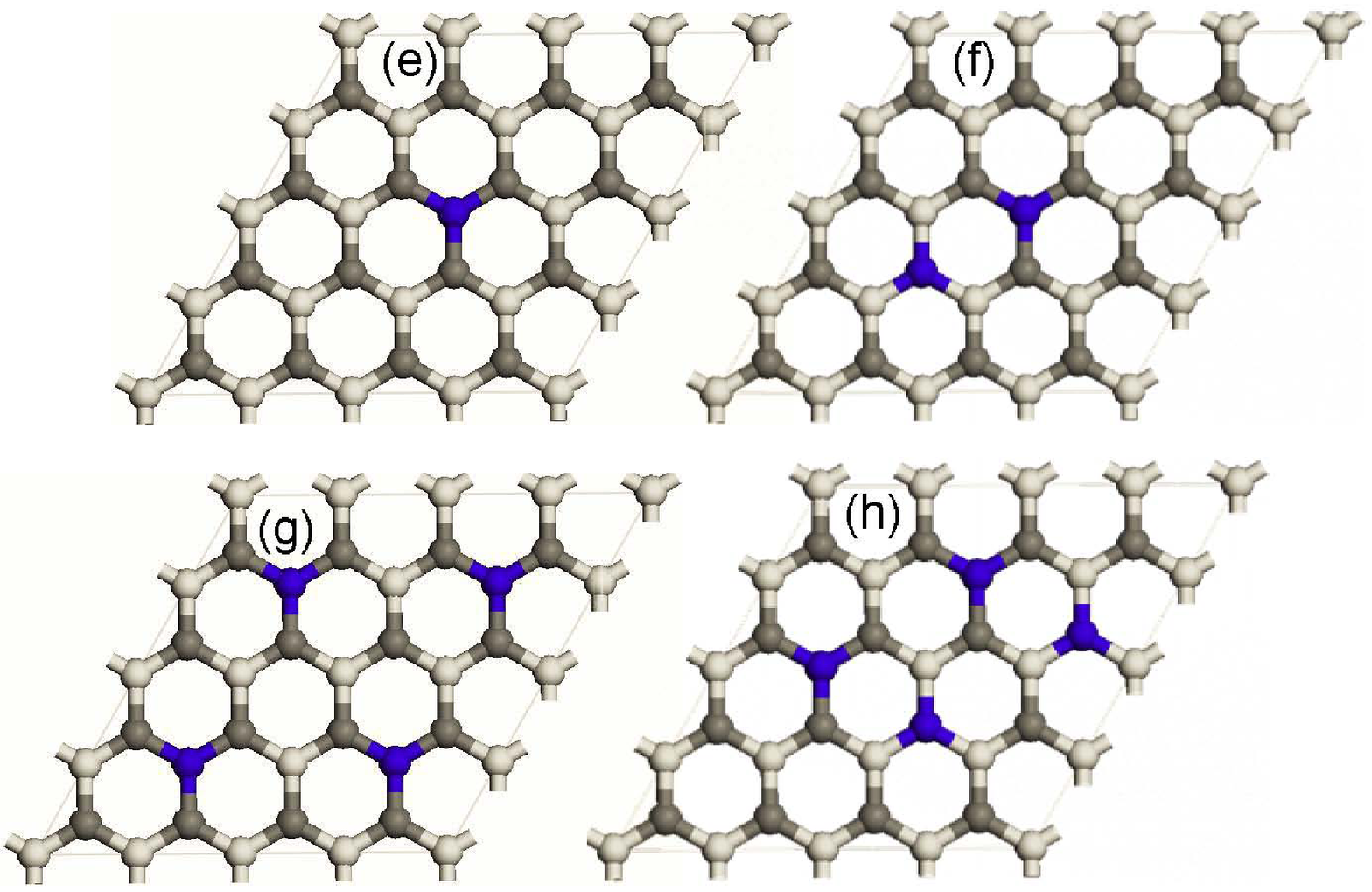}
\includegraphics[width=4.0in]{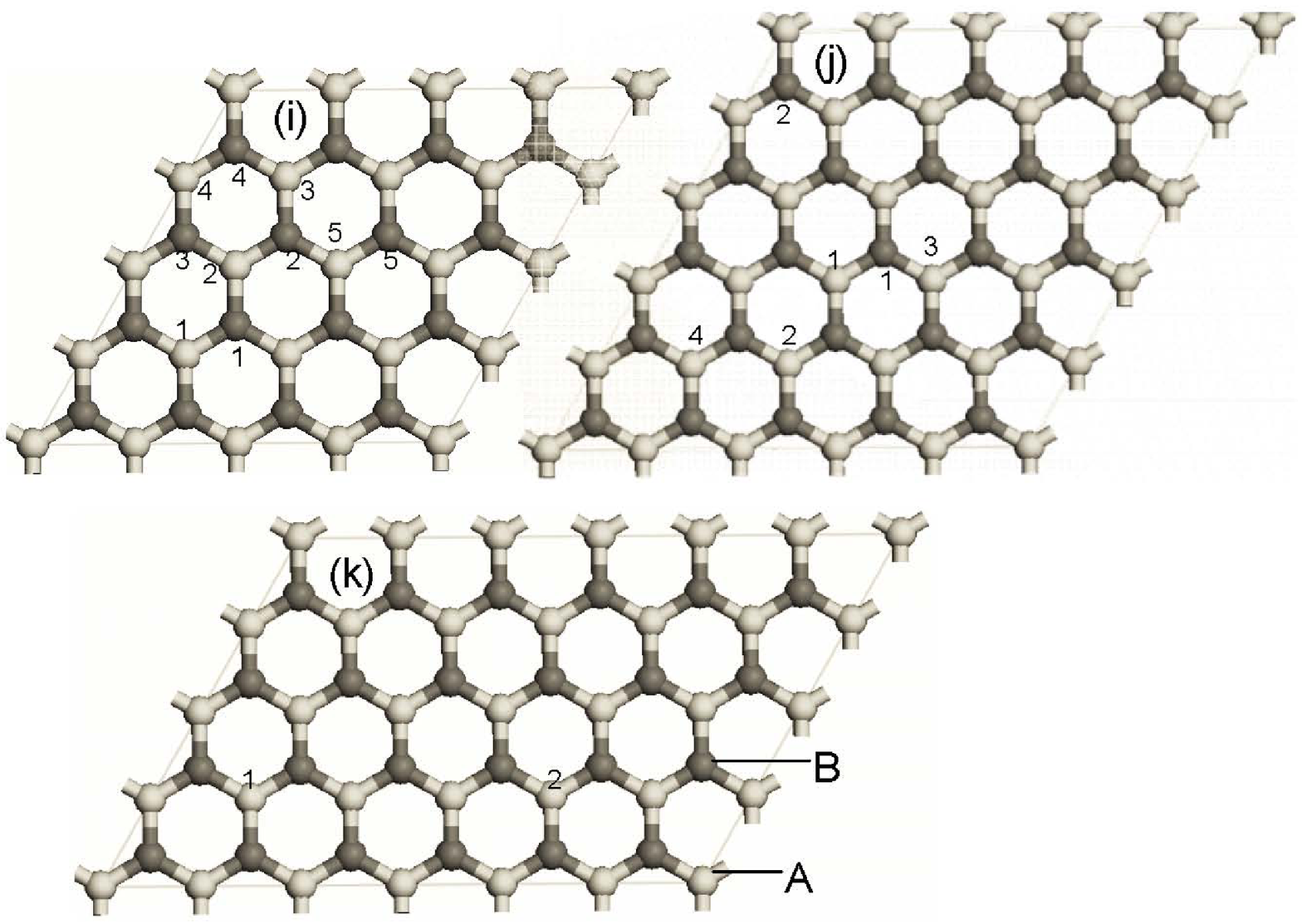}
\renewcommand{\figurename}{Fig.}
\caption{\small{Structures of various graphene sheets with different nitrogen doping configurations,
the dark blue color represents nitrogen atoms.
1(a) and 1(b) sketch one and two nitrogen atoms doped into a 2x2 graphene supercell slab, respectively.
1(c) and 1(d) sketch the similar doping case in a 3x3 graphene supercell slab, respectively.
1(e) and 1(f) sketch the similar doping case too in a 4x4 graphene supercell slab, respectively.
In Fig. 1(g), 4 supercells are displayed, each of which was shown in 1(a).
1(h) presents two couples of para nitrogen atoms doped into a 4x4 graphene supercell slab.
1(i-k) displays three different configurations: 1(i) two sets of potential substitutional doping sites in a 4x4 graphene supercell slab,
1(j) substitutional doping sites in a 5x5 graphene supercell slab, and
1(k) two nitrogens on the same lattice in a 6x4 graphene supercell slab}.}
\label{fig:figure1}
\end{figure}

We can construct a graphene sheet by periodically duplicating a supercell slab in both directions.
Three different types of graphene sheet were calculated, i.e. a 2x2 supercell slab with 8 atoms,
a 3x3 supercell slab with 18 atoms, and a 4x4 supercell slab with 32 atoms.
In order to calculate a two dimensional graphene by the VASP which is basically designed
for three dimensional crystal calculations, we intentionally enlarged the vertical distance
between graphene sheets to 10(\AA) of vacuum, so that the interaction between sheets
reduced to a negligible value. The total energy is expressed as a function of the
unit-cell volume around the equilibrium cell volume $V_{0}$ from experimental data.
The calculated total energies were fitted to the Birch-Murnaghan equation of state to obtain
the optimized lattice constant and a value of 2.469{\AA} was obtained. The structure of the supercells is
shown in Fig.~\ref{fig:figure1}.

Doping effect on the structures and the properties of graphene by a nitrogen mainly
comes from the interactions between the doping atoms and its neighbors.
The formation energy of N doping $E_{f}$ is defined by,
\begin{eqnarray}
E_{f}=E_{dop}-E_{pure}+n(E_{\mu_{C}}-E_{\mu_{N}})
\label{eq:ef0}
\end{eqnarray}
where, $E_{dop(pure)}$ is the energy of doping(pure) system. $E_{\mu}$ is the chemical
potential of C or N atom, and $n$ is the substitutional number.
The atomic energy of C and N atom is -1.316 and -3.114eV, respectively. The difference
between $E_{\mu_{C}}$ and $E_{\mu_{N}}$ equals to that between atomic energy of C and N atom.
The value of $E_{f}$ tells the difficulty to form a doped system.
The smaller the value, the more stable the system.

To make sure the precision of the HSE complication, we compared our calculation results
of the band gaps for 2x2 graphene with two nitrogen atoms adopted to the results which
were reported\cite{Xiang2012}, and we found that they were in excellent agreement.
To assure the accuracy of our calculations, we also reproduced the results of
a doped graphene system, one carbon atom was substituted by a nitrogen atom in 4x6supercell by
other authors. The magnetism, the formation energy and the band structures were in good agreement with
Reference ~\mycite{Singh2009}, if using the same parameters. However, if we used more
k points to get more precise results, the magnetism was not appeared, and correspondingly, larger
Gaussian smearing destroyed the magnetic moment. Therefore, the magnetism had a dependence of temperature
and k-points sampling\cite{Boukhvalov2012}.
\section{Results and discussion}
\subsection{Electronic properties of intrinsic graphene}
\begin{figure}
\center
\setlength{\abovecaptionskip}{-10pt}
\setlength{\belowcaptionskip}{0pt}
\includegraphics[width=3.2in]{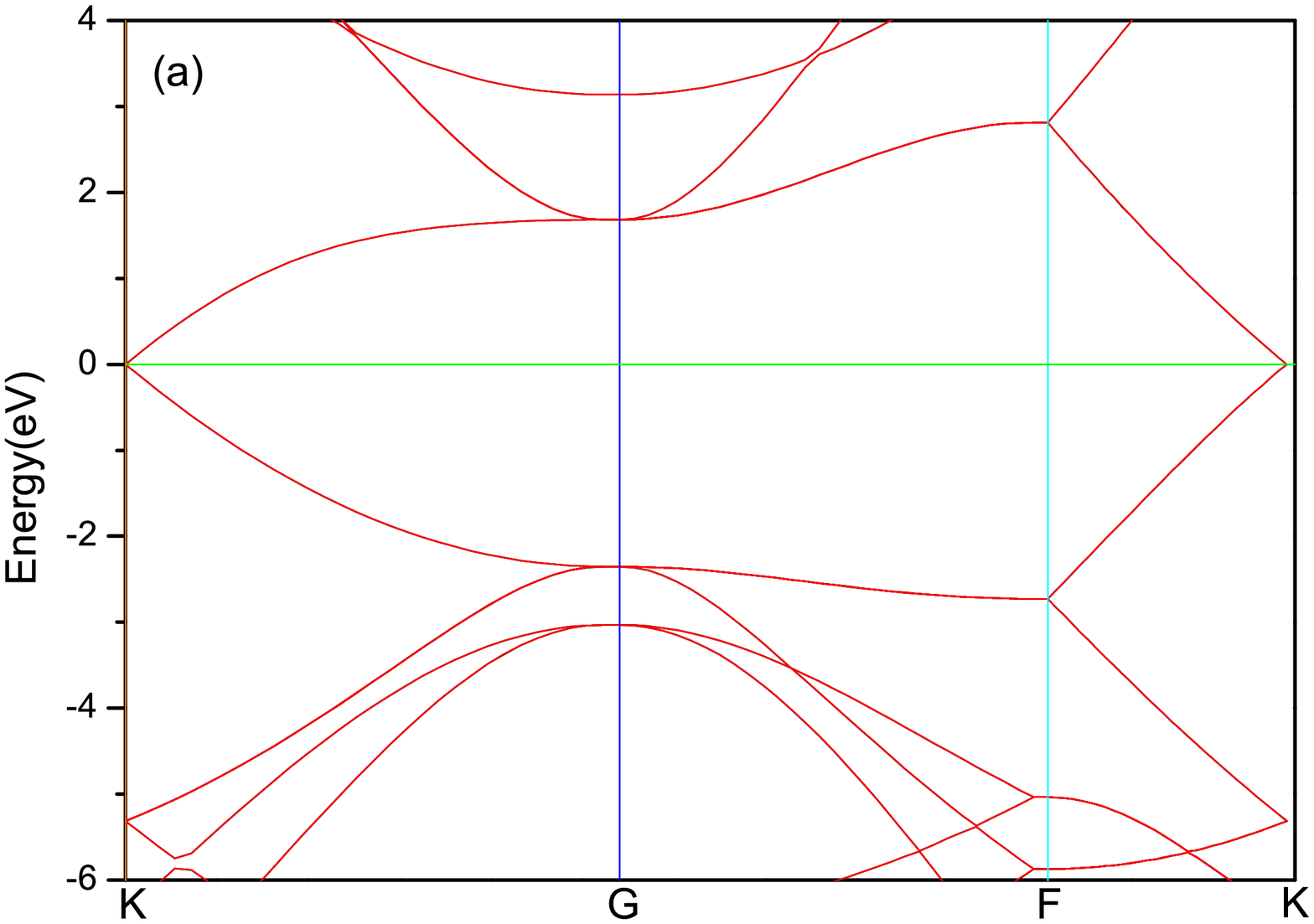}
\includegraphics[width=3.2in]{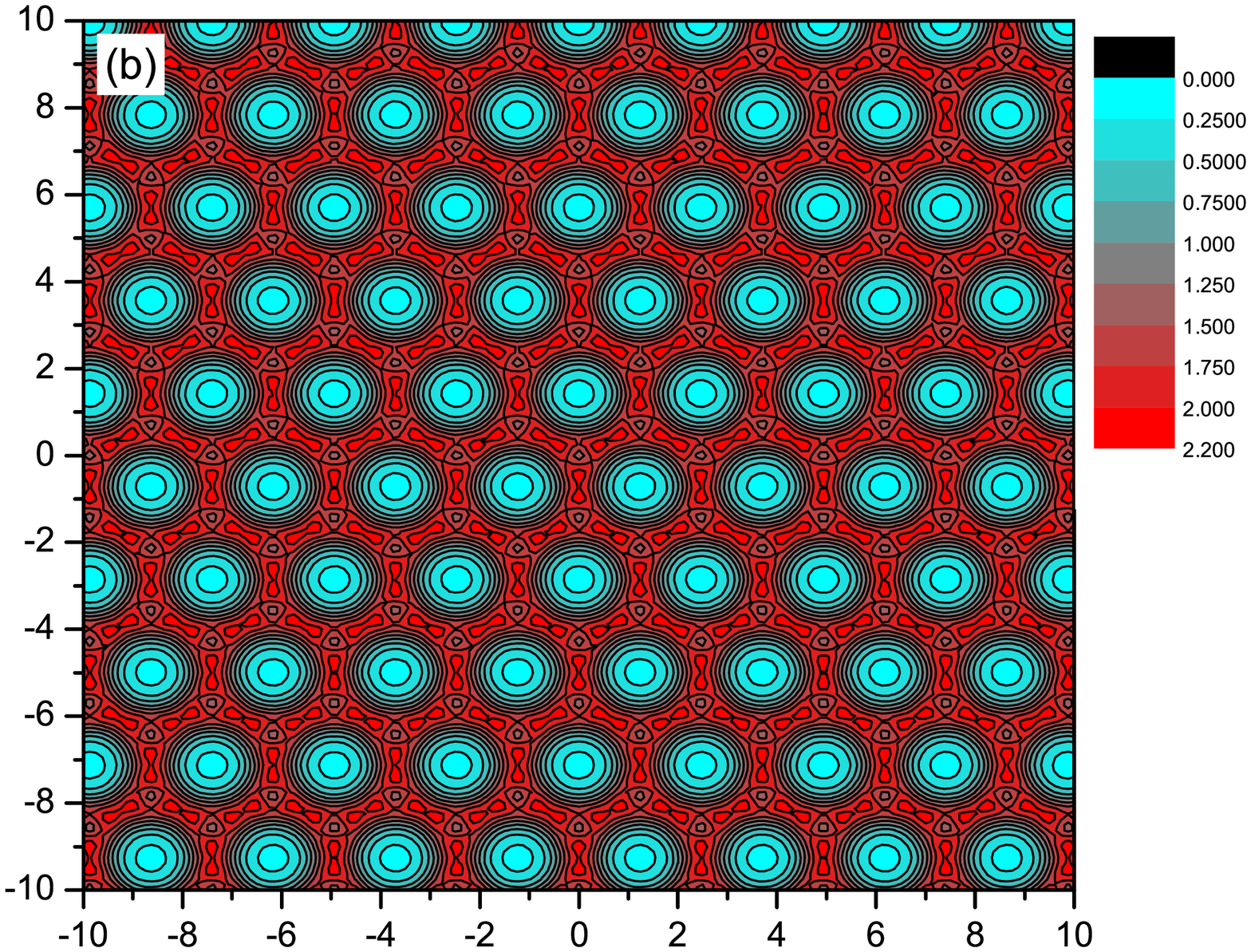}
\renewcommand{\figurename}{Fig.}
\caption{\small{2(a) sketches the band structures of intrinsic 2x2graphene, 2(b) shows the total electron density($\rho_{\alpha}+\rho_{\beta}$)
of pure 2x2graphene, and the plane is 20 by 20 {\AA} across.}}
\label{fig:figure2}
\end{figure}

The calculated electronic properties of pure 2x2graphene are shown in Fig.~\ref{fig:figure2}(a)
and \ref{fig:figure2}(d). The electron distribution illustrates that, the $sp^{2}$ hybridization between two \emph{s} and
two \emph{p} orbitals leads to a trigonal planar structure forming $\sigma$ bonds
between carbon atoms and a $p$ orbital, which is perpendicular to the planar structure,
and interacts covalently with neighboring carbon atoms, forming $\pi$ bonds.
It's well known that half-filled bands in transition elements have played significant role
in the physical characters of correlated systems, the strong interactions lead to a series
of effects, including the magnetism, metal-semiconductor-insulator behaviors, etc.

The band structure shows that graphene is a semiconductor material with zero band gap.
The Fermi level is set to zero energy, and the K point is
the Dirac point of graphene. Graphene is nonmagnetic due to the symmetry between
spin up and down states. The nonmagnetic, zero gap, and massless are three of
the most interesting features of graphene. So comparing
the differences between intrinsic and doped graphene, as well as
confirming the dependency of electronic properties on doping concentrations, are
also meaningful features we should focus on.
\subsection{N-graphene with one nitrogen doping per supercell}
We started with doping cases of one nitrogen atom per graphene supercell, where
one carbon atom was substituted by a nitrogen atom at different supercell
sizes(Fig.~\ref{fig:figure1}(a),\ref{fig:figure1}(c), and \ref{fig:figure1}(e)).
Nitrogen impurities in $\pi$-conjugated systems can in principle affect the magnetic
properties of the material\cite{Hagiri2004}. While in our calculations, only some
configurations showed spin polarization. The magnetic
moments and formation energies of these three doped graphene systems are illustrated in
the Table~\ref{table:table1}. We found that only the 2x2 supercell with one nitrogen
doping had apparent magnetism.
It came from the interactions between itinerant $\pi$ electrons.
The results for smaller doping concentrations we studied were similar to those stated in Reference
~\mycite{Ma2005}, that the ground state of substitutional nitrogen atoms with smaller doping
concentrations was always nonmagnetic even when beginning with an initial magnetic guess configuration.
The magnetism obtained from the HSE calculations was bigger than directly
obtained from GGA or LDA due to HSE calculations employed a more correlated potential.
From the calculated $E_{f}$ values, it can be found that the systems
in which a carbon atom was substituted by a nitrogen atom in smaller concentration were easier
to form, because broking had smaller effect on the system along with the decrease of concentration.
\begin{table}
\center
\caption{\small{The doping concentrations, formation energies ($E_{f}$), magnetic moments
of doping systems with one nitrogen atom, C-N distances, and electrons of nitrogen atoms
for various supercells with one nitrogen doped}.}
\begin{tabular}{|c|c|c|c|}
  \hline
  systems & $~~~$2x2$~~~$ & $~~~$3x3 $~~~$& $~~~$4x4$~~~$ \\\hline
  Doping concentration(at.\%) & 12.5 & 5.56 & 3.125 \\\hline
  $E_{f}$($eV\cdot atom^{-1}$) & 3.852 & 3.724 & 3.636 \\\hline
  magnetic moment per supercell ($\mu_{B}$)  & 0.608 & 0.06 & 0 \\\hline
  C-N distance({\AA}) & 1.426 & 1.418 & 1.414 \\\hline
  Electronic charges of Nitrogen(e) & 6.323 & 6.328 & 6.267 \\
  \hline
\end{tabular}
\label{table:table1}
\end{table}

\begin{figure}
\center
\setlength{\abovecaptionskip}{-10pt}
\setlength{\belowcaptionskip}{0pt}
\includegraphics[width=6in]{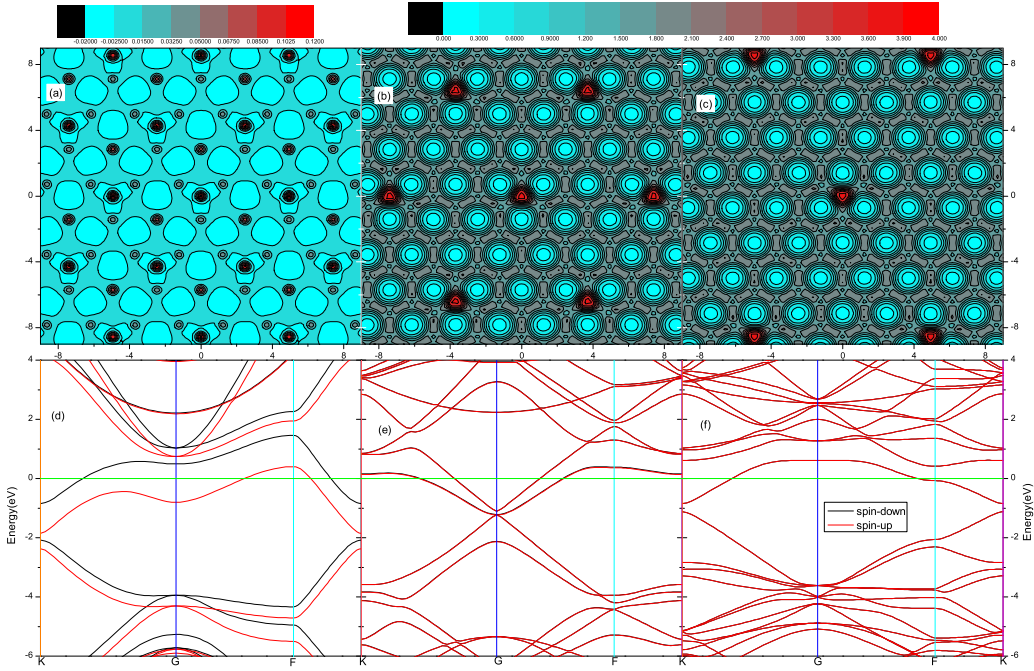}
\renewcommand{\figurename}{Fig.}
\caption{\small{Spin/total electron densities and band structures of N-doped graphene systems
with different concentrations. 3(a) and 3(d) for 2x2 graphene supercell, 3(b) and 3(e) for 3x3 graphene supercell,
3(c) and 3(f) for the 4x4 graphene supercell.
Particularly, 3(a) describes the spin electron densities($\rho_{\alpha}-\rho_{\beta}$) on the (001) plane(i.e., the graphene sheet surface),
3(b) and 3(c) present the total electron densities($\rho_{\alpha}+\rho_{\beta}$), and the planes of 3(a)-(c) are 18 by 18 {\AA} across.
3(d)-3(f) are the band structures of different doping systems with the same energy scales from -6$eV$ to 4$eV$.}}
\label{fig:figure3}
\end{figure}

To visualize the distribution of spin on the graphene sheet, the spin density in 2x2 periodic units
on the graphene sheets is plotted in Fig.~\ref{fig:figure3}(a), and the nitrogen atom is at (0,0).
This clearly indicates that the spin polarization mainly comes from the N atoms,
and the C atoms which are located at the sublattice other than that the N atoms are sited at, as well.
Other carbon atoms carry very small magnetic moments.
Conventionally, magnetism is associated with \emph{d-} or \emph{f-}electrons,
but, the calculated results in our case show that the magnetic moment
is generated by (unpaired) $2p$ electrons of nitrogen atoms.
The bonding status has been changed accordingly, and
the internuclear length of the C-C bonds around the nitrogen atom has also been modified.
The distances between C and N atoms for different supercells are summarized
in Table~\ref{table:table1}, where the original distance between atoms is 1.425{\AA}.
The length of C-N bonds is about the same as that of the original C-C bonds,
so are the bond angles. A Bader analysis on charge number of each atom in the system was performed
to qualify the charge transfers. The results of the nitrogen atom charges in three supercells
are presented in Table~\ref{table:table1}. The electron number being greater than 5e
for N atom suggested that partial electron charges were transferred from the neighbor
C atoms to the more electronegative N atoms. Nitrogen doping results in increasing
positive charge on a carbon atom adjacent to nitrogen atoms and in positive shifting
of Fermi energy at the apex of the Brillouin zone of graphene. As indicated by
Reference ~\mycite{Panchakarla2009}, the fermi level is shifted 0.9 $eV$ in 3.125at.\% N-doped
system. So the charge distributions became nonuniform, and the electrons were gathered
around the N atoms which is the possible source of spin density symmetry breaking.
The breaking is especially visible in the case of 2x2 supercell.

All of these doped systems with one nitrogen per supercell exhibited n-type doping metallic
electronic properties, the band structures of these doped systems were depicted in
Fig.~\ref{fig:figure3}(d)-\ref{fig:figure3}(f). In Fig.~\ref{fig:figure3}(d), we found one majority
spin band was located inside the interval of the minority spin band and crossed the fermi level,
which indicated the strong spin polarization. The majority spin band mainly came from the
coupling interactions between the doping atoms and the carbon atoms in the ortho and para positions of nitrogen
occupied sites. It gave the magnetic moment of 0.608$\mu_{B}$ per supercell.
From \ref{fig:figure3}(e) to \ref{fig:figure3}(f), the spin polarization became smaller, and in some case
disappeared (which can be neglected), i.e., the magnetic moment was significant only in the case that one
nitrogen doped into 2x2 graphene supercell. Table~\ref{table:table1} listed various values of the magnetic
moments for different graphene supercells. And Table~\ref{table:table2} listed the atomic magnetic values for
single nitrogen atom in 2x2 supercell, the values characterized the itinerant properties of $\pi$ electrons.
\begin{table}
\center
\caption{\small{The atomic magnetic values for 2x2 supercell with one nitrogen atom doped}.}
\begin{tabular}{|c|c|c|c|c|}
  \hline
  atom on A sublattice         & $C_{A_{1}}$ & $C_{A_{2}}$ & $C_{A_{3}}$ & $N$ \\\hline
  magnetic moment ($\mu_{B}$)  &  0.012      &  0.010      &  0.011      & -0.146 \\\hline
  atom on B sublattice         & $C_{B_{1}}$ & $C_{B_{2}}$ & $C_{B_{3}}$ & $C_{B_{4}}$ \\\hline
  magnetic moment ($\mu_{B}$)  & -0.193      & -0.101      & -0.101      & -0.100    \\
  \hline
\end{tabular}
\label{table:table2}
\end{table}
\subsection{N-graphene with two nitrogen  doping per supercell}
We also studied the two nitrogen doping case in one supercell, by using the same methods.
Two carbon atoms in a six-number ring were replaced by two nitrogen atom. Comparing the
formation energy of doping atoms at three different positions in a ring, i.e. ortho-,
meta-, and para-positions, the case with two nitrogen atoms in a para position had the
smallest formation energy and had the most stable structures(illustrated in the Table ~\ref{table:table3}).
This was in good agreement with the reported results that the interaction energy curve had
a local minima when two nitrogen atoms in a para position\cite{Xiang2012}.
We calculated the three different cases, where the carbon atoms
in a ring were substituted by two nitrogen atoms
in 2x2, 3x3, and 4x4 graphene supercells respectively.
The substitutional positions were described in Fig.~\ref{fig:figure1}(b), \ref{fig:figure1}(d) and \ref{fig:figure1}(f)
for 2x2, 3x3 and 4x4supercells. We reproduced the band structure calculation for the 2x2 graphene supercell
with two nitrogen atoms doping, which was reported by Reference ~\mycite{Xiang2012}, and obtained the similar
results. The formation energies and Bader electrons for the nitrogen atom for these three
doped graphene systems are listed in Table~\ref{table:table3}.
The average electrons that the nitrogen atoms possessed were greater than that
of one nitrogen doping, more charges had been transferred.
\begin{table}
\center
\caption{\small{The doping concentrations, formation energies ($E_{f}$), magnetic moments of doping systems with
two nitrogen atom, C-N distances, and electrons of nitrogen atoms for various supercells with two nitrogen atoms doped}.}
\begin{tabular}{|c|c|c|c|}
  \hline
  system & 2x2 & 3x3 & 4x4 \\\hline
  Doping concentration(at.\%) & 25 & 11.11 & 6.25 \\\hline
  $E_{f}$($eV\cdot atom^{-1}$) & 3.627 & 3.625 & 3.674 \\\hline
  magnetic moment($\mu_{B}$) & 0 & 0 & 0 \\\hline
  C-N distance({\AA}) & 1.427 & 1.418/1.425 & 1.414/1.422 \\\hline
  Electronic charges of Nitrogen(e) & 6.284/6.355 & 6.287/6.358 & 6.297/6.367 \\
  \hline
\end{tabular}
\label{table:table3}
\end{table}

\begin{figure}
\center
\setlength{\abovecaptionskip}{-10pt}
\setlength{\belowcaptionskip}{0pt}
\includegraphics[width=6in]{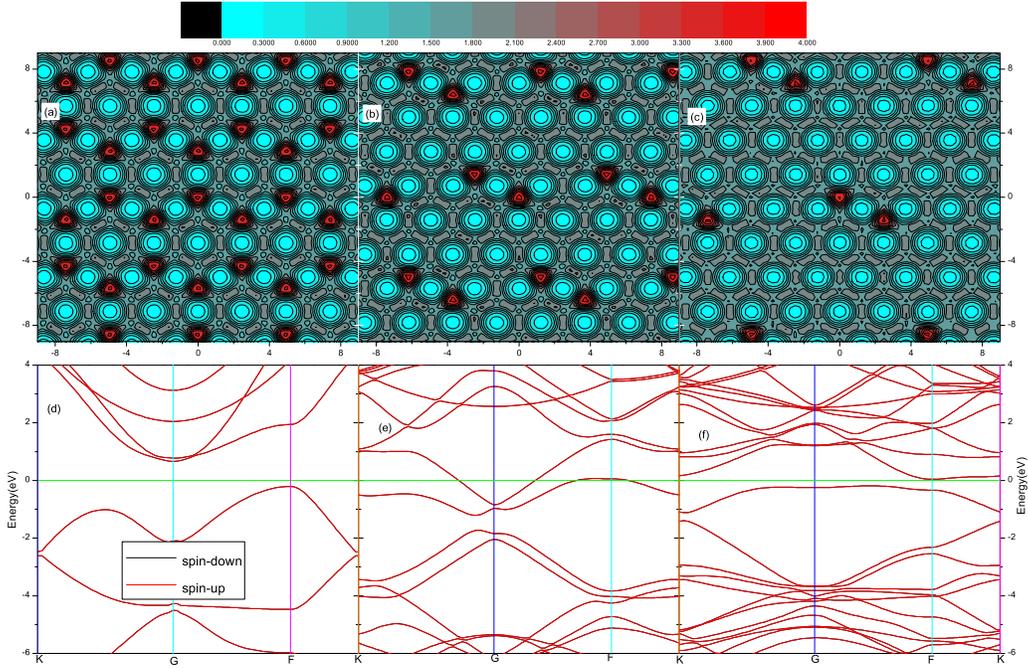}
\renewcommand{\figurename}{Fig.}
\caption{\small{(Total($\rho_{\alpha}+\rho_{\beta}$) electron densities and band structures of
doped graphene systems with two nitrogen atoms doped into different units, where
4(a) and 4(d) for 2x2 graphene sheet, 4(b) and 4(e) for 3x3 graphene sheet, 4(c) and 4(f) for the 4x4 graphene sheet.
Particularly, 4(a), 4(b), and 4(c) present the total electron densities on the (001) plane(i.e., the graphene sheet surface)
and all of them have the same scales. The planes for them are 18 by 18 {\AA} across.
4(d)-4(f) describe the band structures of different doping systems with the same energy scales from -6eV to 4eV}.}
\label{fig:figure4}
\end{figure}

Calculated results for different configurations were presented in Fig.~\ref{fig:figure4}.
There were no spin polarizations in any sizes of supercells with two nitrogen atoms doping,
so we plotted the total electron densities of these three supercells in Fig.~\ref{fig:figure4}(a)-\ref{fig:figure4}(c)
to investigate the bonding effects after nitrogens were doped into the supercells in comparison
with the intrinsic graphene in Fig.~\ref{fig:figure2}(a). We can see from the figures, that
the electron density was increased from Fig.\ref{fig:figure4}(a) through Fig.\ref{fig:figure4}(c),
and more electrons were gathered in the vicinity of a nitrogen atom. The three carbon atoms
which were closed to a nitrogen atom are located in the same isosurface. This is the evidence that
the original $C=C$ bonds had been broken to form new C-N bonds in the hexagonal structure.
There are no magnetic moment existed in any local atoms. We also can see that Fig.\ref{fig:figure4}(a)
gives a perfect continuously symmetric six-ring.

Both of 2x2 and 4x4 graphene supercell sheet with two nitrogen doping atoms have exhibited
n-type semiconductor behavior with a small band gap. But a 3x3 graphene supercell shows metal
electronic properties whose band structures were depicted in Fig.~\ref{fig:figure4}(d)-\ref{fig:figure4}(f).
All of \ref{fig:figure4}(d)-\ref{fig:figure4}(f) showed symmetry between majority and minority spin.
The band gaps were 1.0 and 0.23eV for 2x2 and 4x4 graphene supercell with two nitrogen doping atoms, respectively,
in agreement with the existing value of 0.96eV in 2x2 sheet\cite{Xiang2012}. Reference ~\mycite{Xiang2012}
had explained the band gap was produced because the hopping between N $p_{z}$ orbitals in the configurations that
led to the larger dispersion of the highest occupied band. Doping mechanism will make a system disorder.
And we know that ordered structures are more stable in lower temperatures. Symmetric doping may
lead to a stable system. From our calculations, we can suspect that a graphene supercell with
two para nitrogen atoms in a ring as well as with multiplicity of two para nitrogen atoms
in a ring would not be spin polarized. The units which are the $n\times n$ ($n$ is not divided by 3)
with para nitrogen doping will produce a band gaps.
\subsection{N-graphene with four nitrogen atoms doping per supercell}
From the above discussion, we learnt that electronic structures were differently depend on
the different concentrations and the species. The next question is whether systems that have the same
doping concentration(12.5\%) but different configurations have the
same results. Two configurations with four nitrogen atoms doped into
4x4graphene sheets were investigated and compared with that of one nitrogen atom doped
into 2x2 graphene sheet. The one case is four carbon atom sites in different ring
were substituted by nitrogen atoms(shown in Fig.~\ref{fig:figure1}(h)). The formation energy of doping
for 4x4 graphene sheet with four nitrogen atoms is 3.731eV/atom from Eq.~(\ref{eq:ef0}),
suggesting that the formation energy of N-doping has no direct relation with concentrations.
It has a small magnetic moment due to the coupling between the closer nitrogen atoms in
two pairs of para positions. And it has a small gap.
The other one is four times 2x2 graphene with one nitrogen doped(shown in Fig.~\ref{fig:figure1}(g)).
The formation energy(3.845eV/atom) of doping and band structures were the same with
that of 2x2 graphene with one nitrogen doped. And the magnetic
moment is 2.5$\mu_{B}$, nearly four times that of 2x2 graphene with one nitrogen.
These results from the two substitutes indicate
that the magnetic and electronic properties are not only relied on the concentration but also relied on the
configurations of the nitrogen arranged.

\subsection{Exploring the origin of the magnetism}
To reveal the origin of the magnetism, we calculated various structures differed by doping positions
of two nitrogen atoms in a 4x4 graphene supercell, which were shown in Fig.~\ref{fig:figure1}(i).
A and B were labeled in aqua and grey colors, respectively, representing substituted sites in
different sublattices for the honeycomb lattice. For the 4x4 graphene supercell, the largest
distance between two doping nitrogen atoms is $\frac{4\sqrt{3}}{3}a < 3a$, where $a$ is the lattice constant
for the unit cell. The magnetic moments, doping formation energy, along with the distance between two doping
atoms are shown in Table~\ref{table:table4}.
\begin{table}
\center
\caption{\small{The formation energies ($E_{f}$), magnetic moments of doping systems doped by
two nitrogen atoms for 4x4 graphene sheet with increasing N-N distances}.}
\begin{tabular}{|c|c|c|c|}
  \hline
  distance & sublattice site & $E_{f}$($eV\cdot atom^{-1}$) & magnetic moment($\mu_{B}$)\\\hline
  $\frac{\sqrt{3}}{3}$a  & $A_{1}B_{1}$ &  4.174 & 0 \\\hline
  a                      & $A_{1}A_{2}$ &  3.837 & 0.75 \\\hline
  $\frac{2\sqrt{3}}{3}$a & $A_{1}B_{3}$ &  3.674 & 0 \\\hline
  $\frac{\sqrt{21}}{3}$a & $A_{1}B_{2}$ &  3.744 & 0 \\\hline
  $\sqrt{3}$a            & $A_{1}A_{4}$ &  3.734 & 0.125 \\\hline
  2a                     & $A_{1}A_{3}$ &  3.759 & 0.625 \\\hline
  $\frac{\sqrt{39}}{3}$a & $A_{1}B_{4}$ &  3.627 & 0 \\\hline
  $\frac{4\sqrt{3}}{3}$a & $A_{1}B_{5}$ &  3.657 & 0 \\\hline
\end{tabular}
\label{table:table4}
\end{table}
From the table, we found three interesting phenomena in a 4x4 graphene supercell.
The first, when two nitrogen atoms replaced two carbon atoms along the zigzag direction
in the same sublattice, and the distance between two doping atoms is $\leq$ 2a,
there will be an approximate 0.7$\mu_{B}$ magnetic moment.
The second, when the doping was occurred in the armchair direction,
there would be negligible even no magnetic moment.
The third, when two carbon atoms in different sublattices were doped by nitrogen atoms,
there would be no spin polarization.

By analyzing the two magnetic conditions $A_{1}A_{2}$ and $A_{1}A_{3}$, we found both of them
produce some nearly zero-magnetism sublattice lied between them. Other carbon atoms in B lattice
show FM coupling with N atoms, whereas the carbon atoms in the same sublattice with N atoms
show AFM coupling. This phenomenon  satisfies the RKKY-like coupling in graphene for
on-site impurities\cite{Black-Schaffer2010}. And the magnetic moments on the two nitrogen atoms for
the former are larger than that for the later. From the forming energies plotted in various distance of
dopants, there are two minimum values for $A_{1}B_{3}$ and $A_{1}B_{4}$ cases. The configurations
agree with that reported by others in 12x12 supercell\cite{Xiang2012}.

In order to obtain large relative distance between two nitrogen atoms, we use 5x5 and 4x6 supercells,
which were shown in Fig.~\ref{fig:figure1}(j) and (k), to calculate the dependence of magnetism on
distances of two impurities sites. Four different configurations($A_{1}B_{1}$,$A_{1}A_{3}$,
$A_{3}A_{4}$,and $A_{2}B_{2}$) in a 5x5 graphene sheet and one($A_{1}A_{2}$) in a 4x6 graphene are calculated.
The results are shown in Table~\ref{table:table5}.
\begin{table}
\center
\caption{\small{The formation energies ($E_{f}$), magnetic moments of doping systems doped by
two nitrogen atoms for 5x5 and 6x4 graphene sheet with increasing N-N distances}.}
\begin{tabular}{|c|c|c|c|c|}
  \hline
  system & distance & sublattice site & $E_{f}$($eV\cdot atom^{-1}$) & magnetic moment($\mu_{B}$)\\\hline
  5x5 & $\frac{\sqrt{3}}{3}$a & $A_{1}B_{1}$ &  4.172 & 0 \\\hline
  5x5 & a & $A_{1}A_{3}$ & 3.805 & 0 \\\hline
  5x5 & $\sqrt{7}$a & $A_{3}A_{4}$ & 3.67 & 0 \\\hline
  5x5 & $\frac{5\sqrt{3}}{3}$a & $A_{2}B_{2}$ & 3.515 & 0 \\\hline
  6x4 & 3a & $A_{1}A_{2}$ & 3.685 & 0 \\\hline
\end{tabular}
\label{table:table5}
\end{table}
From Table~\ref{table:table4}, we can infer the magnetism vanishes with the
increasing distance between two doping nitrogen atoms in any direction and sublattice.
The two same case(orth and meta positions) for 4x4 and 5x5, a little small doping
forming energy are obtained for 5x5 system, since it has smaller doping concentration.
Two nitrogen atoms substitute two same lattice along zigzag direction with 3a distance
in a 6x4 graphene has zero magnetic moment. Therefore, the magnetism vanishes with the
decreasing of doping concentration, and the increasing of the distance between two nitrogen atoms.

It's important to note that, we have reproduced the calculation that one carbon atom was substituted
by one nitrogen atom for 4x6supercell, if we use smaller k-points which used in Reference ~\mycite{Singh2009},
the magnetic moment was 0.556$\mu_{B}$, similar to the reference result of 0.46$\mu_{B}$. The difference stems from
the HSE and DFT calculation stated in Reference ~\mycite{Marsman2008}. But, if we increase the number of
k-mesh for more precise results, no any magnetism has been found.

The above analyses strongly implied that the sublattice symmetry is crucial to induction of
magnetic moments. The sublattice symmetry breaking will possibly break energy degeneracy with
respect to spin (spin symmetry breaking), which is the origin of spin-polarization,
and the occurrence of magnetic response.

\section{Conclusion}
Making use of the first-principles density functional theory (DFT) based methods,
we calculated N-graphene properties in various doping concentrations and configurations.
The calculated results showed that the graphene with one nitrogen doped in any size
of graphene supercells, the graphene with two nitrogen atoms doped in a 3x3 graphene supercell
exhibited n-type metal behavior, and the graphene with two nitrogen atoms doped into a 2x2
and a 4x4supercell had band gaps of 1.0 and 0.23eV, respectively. Especially, for a 2x2
graphene system with two nitrogen atoms doped, had a similar band structure as silicon.
Four nitrogen atoms(two para) doped into a 4x4 graphene also opened band gaps. From these
features, we proposed a way to tailor band gaps according to the
size of graphene unit and atom configurations.

The systems with single nitrogen atom doped in a 2x2graphene sheet displayed spin polarizations,
and spin couplings were disappeared for some special configurations. We also found graphene
with two nitrogen doping atoms at para positions in a ring did not have spin polarizations
for any sizes of graphene supercells. From our analysis of magnetism, interactions between
nitrogen atoms and carbon atoms led to the magnetism, and the magnetism was closely
related to the sublattice symmetry and distance between nitrogen atoms. We should point
out sublattice symmetry is significantly important, which can be seen from the comparisons
between a 4x4 graphene sheet with two couples of para nitrogen atoms and one nitrogen
doped into 2x2 graphene sheet, as well as the conditions listed in the Table~\ref{table:table4}.

Our results provided a comprehensive analysis about N-doped graphene, which should be a hint for producing graphene-based devices
and other multiple applications as well.
\bibliography{library}
\end{document}